# Utilizing the information content in two-state trajectories

Ophir Flomenbom & Robert J. Silbey[*]

*Department of Chemistry, Massachusetts Institute of Technology, Cambridge, MA 02139*

**The signal from many single molecule experiments monitoring molecular processes, such as enzyme turnover via fluorescence and opening and closing of ion-channel via the flux of ions, consists of a time series of stochastic *on* and *off* (or open and closed) periods, termed a two-state trajectory. This signal reflects the dynamics in the underlying multi-substate *on-off* kinetic scheme (KS) of the process. The determination of the underlying KS is difficult and sometimes even impossible due to the loss of information in the mapping of the mutli-dimensional KS onto two dimensions. Here we introduce a new procedure that efficiently and optimally relates the signal to all equivalent underlying KSs. This procedure partitions the space of KSs into canonical (unique) forms that can handle *any* KS, and obtains the topology and other details of the canonical form from the data without the need for fitting. Also established are relationships between the data and the topology of the canonical form to the *on-off* connectivity of a KS. The suggested canonical forms constitute a powerful tool in discriminating between KSs. Based on our approach, the upper bound on the information content in two-state trajectories is determined.**

The data from a wide range of single molecule experiments (1-23), i.e. the passage of ions and biopolymers through individual channels (3-5), activity and conformational changes of biopolymers (6-15), diffusion of molecules (16-19), and blinking of nano-crystals (20-23), is inevitably turned into a trajectory of *on* and *off* periods (waiting times), FIG 1A. A frequently used assumption describes the mechanism of the observed process by a multi-substate *on-off* Markovian kinetic scheme (KS) (24-36), Fig. 1B. (This is a fairly unrestrictive assumption because, in many cases, adding substates to the KS is equivalent for describing the process by coupled stochastic (sub-) processes; see also Refs. 37-50). The KS describes a discrete conformational energy landscape of a biomolecule, chemical kinetics with (or without) conformational, or environmental, changes, stands for quantum states, etc. The underlying stochastic dynamics of the process in the multi-substate *on-off* KS is thus encoded in the two-state trajectory (the stochastic signal changes value only when transitions between substates of different states in the KS take place). The aim of such sophisticated single molecule experiments is to learn about the underlying KS to an extent that is unattainable from bulk measurements due to averaging. However, determining the KS from the two-state trajectory is difficult since the number of the substates in each of the states, $L_x$ ($x = on, off$), is usually large, and the connectivity among the substates is generally complex. A widely used approach for deducing the KS relies on the construction of waiting time probability density functions (WT-PDFs): the WT-PDF of state $x$ (= $on, off$), $\phi_x(t)$, and the joint PDFs of two successive waiting times, $x$ event followed by $y$ event, $\phi_{x,y}(t_1,t_2)$, $x, y = on, off$. [Higher order successive WT-PDFs do not contain additional information on top of $\phi_{x,y}(t_1,t_2)$ (33)]. $\phi_x(t)$ and $\phi_{x,y}(t_1,t_2)$ are fitted to sums of exponents by common methods, e.g. Ref. 51. Then, a search for a KS that leads to the fitted PDFs is performed. Alternatively, a maximum likelihood method can be applied (24-25), which demands first assuming a KS's topology. Although these techniques are frequently used, looking for a possible KS that reproduces the data is an exhaustive task. Moreover, there are KSs with the same $\phi_x(t)$s and $\phi_{x,y}(t_1,t_2)$s (26-32). A more sophisticated approach divides the kinetic scheme space into canonical (unique) forms. (Underlying KSs with the same canonical form are equivalent to each other; see, however, the discussion in, *Examples and the utility of RD forms*). Two divisions into canonical forms were previously suggested, called, following Bruno *et al.* (30), MIR [manifest interconductance rank (30)], and BKU [Bauer–Kienker uncoupled (31-32)] forms. MIR and BKU forms are useful in handling reversible connection, non-symmetric (i.e. the spectrums of the $\phi_x(t)$s is non-degenerate), underlying KSs, and are not so efficient in discriminating between KSs. In practice, MIR and BKU forms are found from the data using fitting procedures. Here, we introduce new canonical forms, called reduced dimensions (RD) forms, which can handle *any* KS, i.e. a KS with irreversible connections and/or symmetry. Relationships between the data, the KS's *on-off* connectivity and the RD form's topology are established. These relationships are used in mapping a KS into a RD form. A simple procedure for finding the RD form from the data is given, where the topology and other details of the RD form are determined without the need of fitting, which significantly shortens the search time in the kinetic scheme space. The suggested canonical forms constitute a powerful tool in discriminating among KSs. Based on our approach, the upper bound on the information content in two-state trajectories is set.

**METHODS**

***Explicit on-off connectivity representation of the WT-PDFs*** Our approach is based on expressing the WT-PDFs in an explicit *on-off* connectivity representation (for any KS). As usual, the *on-off* process is separated into two irreversible processes that occur sequentially (24-36). For example, $\phi_{x,y}(t_1,t_2)$ ($x \neq y$) is given by,

$$\phi_{x,y}(t_1,t_2) = \sum_{n_y=1}^{N_y} \left( \sum_{n_x=1}^{N_x} W_{n_x} f_{n_y n_x}(t_1) \right) F_{n_y}(t_2) = \sum_{m_x \in \{M_x\}} \left( \sum_{n_x=1}^{N_x} \sum_{n_y=1}^{N_y} W_{n_x} \tilde{f}_{m_x n_x}(t_1) \omega_{n_y m_x} F_{n_y}(t_2) \right). \quad (1)$$

(In section A of the supporting information (SI.A), which is accompanied the paper on the PNAS website, expressions for $\phi_x(t)$ and $\phi_{x,x}(t_1,t_2)$ are given). Equation (1) emphasizes the role of the KS's topology in expressing the $\phi_{x,y}(t_1,t_2)$s. $N_x$ and $M_x$ are the numbers of initial and final substates in state *x* in the KS, respectively. Namely, each event in state *x* starts at one of the $N_x$ initial substates, labeled, $n_x=1,...,N_x$, and terminates through one of the $M_x$ final substates, labeled $m_x = 1,..., M_x$, for a reversible *on-off* connection KS (Fig. 1B), or $m_x = N_x +1-H_x,..., N_x +M_x-H_x$, for an irreversible *on-off* connection KS (Fig. 2A), where $H_x$ (= 0,1,..., $N_x$) is the number of substates in state *x* that are both initial and final ones. (In each of the states the labeling of the substates starts from 1). An event in state *x* starts in substate $n_x$ with probability $W_{n_x}$. The first passage time PDF for exiting to substate $n_y$, conditional on starting in substate $n_x$ ($x \neq y$), is $f_{n_y n_x}(t)$, and $F_{n_x}(t) = \sum_{n_y} f_{n_y n_x}(t)$. Writing $f_{n_y n_x}(t)$ as, $f_{n_y n_x}(t) = \sum_{m_x} \omega_{n_y m_x} \tilde{f}_{m_x n_x}(t)$, emphasizes the role of the *on-off* connectivity, where $\omega_{n_y m_x}$ is the transition probability from substate $m_x$ to substate $n_y$, and $\tilde{f}_{m_x n_x}(t) \omega_{n_y m_x}$ is the first passage time PDF, conditional on starting in substate $n_x$, for exiting to substate $n_y$ through substate $m_x$. (A sum $z_x \in \{Z_x\}$ is a sum over a particular group of $Z_x$ substates). Note that all the factors in Eq. (1) can be expressed using the master equation (SI.B).

**RESULTS AND DISCUSSION**

***The rank of*** $\phi_{x,y}(t_1, t_2)$ ***and it's topological interpretation*** For discrete time, $\phi_{x,y}(t_1, t_2)$ is a matrix, whose rank $R_{x,y}$ (= 1, 2, …), which is the number of non-zero eigenvalues (or singular values for a non square matrix) of it's decomposition, can be obtained without the need of finding the actual functional form of $\phi_{x,y}(t_1, t_2)$. Using Eq. (1), which gives $\phi_{x,y}(t_1, t_2)$ as sums of terms each of which is a product of a function of $t_1$ and a function of $t_2$, we can relate $R_{x,y}$ ($x \neq y$) to the topology of the underlying KS. When none of the terms in an external sum on Eq. (1), after the first or the second equality, are proportional, $R_{x,y} = \min(M_x, N_y)$ (Fig. 1A). Otherwise, $R_{x,y} < \min(M_x, N_y)$ (Fig. 2E, and SI.C), and Eq. (1) is rewritten such that it has the *minimal* number of additives in the external summations,

$$\phi_{x,y}(t_1, t_2) = \sum_{n_y \in \{\tilde{N}_y\}} \left( \sum_{n_x=1}^{N_x} W_{n_x} f_{n_y n_x}(t_1) \right) F_{n_y}(t_2)$$

$$+ \sum_{m_x \in \{\tilde{M}_x\}} \left( \sum_{n_x=1}^{N_x} W_{n_x} \tilde{f}_{m_x n_x}(t_1) \right) \left( \sum_{n_y \notin \{\tilde{N}_y\}} \omega_{n_y m_x} F_{n_y}(t_2) \right). \quad (2)$$

This leads to the equality, $R_{x,y} = \tilde{N}_y + \tilde{M}_x$. $\tilde{N}_y$ and $\tilde{M}_x$ can be related to the KS's *on-off* connectivity. Consider a case where $M_x < N_y$, and there is a group of final substates in state $x$, $\{O_x\}$, with connections *only* to a group of initial substates in state $y$, $\{O_y\}$, and $O_x > O_y$ (see Fig. C4 in SI.C). Then $\widetilde{M}_x = M_x - O_x$ and $\widetilde{N}_y = O_y$. (Further discussion and a generalization of this relationship are given in SI.C).

***The RD form*** The $R_{x,y}$s are obtained from the $\phi_{x,y}(t_1,t_2)$s without the need of finding its' actual functional forms, thus constitute a fitting-free relationship between the data to the *on-off* connectivity and details of the underlying KS. Utilizing this relationship, the kinetic scheme space is divided into canonical forms, RD forms, using the $R_{x,y}$s. Excluding KSs with symmetry, $R_{x,y}$ ($x \neq y$) is the number of substates in state $y$ in the RD form (see also the discussion in *Additional relationships between the data, the RD form, and the KS*). RD forms can represent underlying KSs with symmetry and irreversible connections because they are built from all four $R_{x,y}$s. RD form has the minimal number of substates needed to reproduce the data. This number is smaller or equal to the number of independent *on-off* connections in the MIR form (SI.D). (The equality holds for non-symmetric, reversible connection, KSs). Connections in the RD form are only between substates of different states, as in the BKU form. Unlike the MIR and BKU forms, for each connection in the RD form there is a WT-PDF that is not necessarily exponential.

***Mapping a KS into a RD form*** Mapping a KS into a RD form is based on clustering of (some of) the initial substates in the KS, depending on the KS's *on-off* connectivity. Such clusters are one of the two kinds of substates in the RD form, where the second kind originates from single initial substates in the KS. For a non-symmetric KS, initial substates in state $y$ in the KS that contribute to $R_{x,y}$ ($x \neq y$) are mapped to themselves and those that do not contribute to $R_{x,y}$ are clustered, where initial-$y$-state substates in a cluster are all connected to the same final-$x$-state substate that contributes to $R_{x,y}$. (When substate $m_x$ has a single exit-connection to substate $n_y$, which is it's only entering-connection, substate $n_y$ is defined as the one contributing to the rank). For example, the KS in Fig. 1B is mapped into a RD form (Fig. 3D) when clustering substates $1_{off}$-$2_{off}$ and substates $3_{off}$-$5_{off}$ into the RD form's substates $1_{off}$ and $2_{off}$, respectively, because non of the initial-*off*-state substates contribute to $R_{on,off}$. The initial *on* substates are mapped to themselves because they both contribute to $R_{off,on}$.

The clustering procedure fully determines the WT-PDFs for the connections in the RD form (technical details for obtaining these WT-PDFs given a KS are discussed in SI.C). Note that the clustering procedure, along with the fact that substates in the KS that are not initial or final ones do not affect the RD form's topology, reduce the KS dimensionality to that of the RD form.

***Finding the RD form from the data*** The following steps can be used for finding the RD form from the data (when fitting is needed, we rely on known procedures, e.g. Refs. 24-25, 51): (*1*) Find the number of substates in the RD form using decomposition of the $\phi_{x,y}(t_1,t_2)$s. (*2*) Obtain the spectrum of the $\phi_x(t)$s using fitting procedures. The spectrum of the WT-PDFs for the *x* to *y* connections in the RD form is the same spectrum as that of $\phi_x(t)$, because substates of the same state in the RD form are not connected. Differences lay in the pre-exponential coefficients. (Steps (*1*)-(*2*) can be permutated). (*3*) Apply fitting procedures for finding the pre-exponential coefficients of the WT-PDFs for the connections in the RD. (Other technical details for constructing the RD form from the data are discussed in SI.E).

***Examples and the utility of RD forms*** The simplest topology for a RD form has one substate in each of the states, namely, $R_{x,y}=1$ ( $x,y=on,off$ ), and the only possible choice for the WT-PDFs for the connections is $\phi_{on}(t)$ and $\phi_{off}(t)$ (Fig. 2D). This means that all the information in the data is contained in $\phi_{on}(t)$ and $\phi_{off}(t)$. Consequently, KSs with $R_{x,y}=1$ ( $x,y=on,off$ ) and the same $\phi_{on}(t)$ and $\phi_{off}(t)$ are indistinguishable (assuming no additional information on the mechanism is known). Examples of such KSs are shown in Fig. 2A-2C. This case was discussed in Refs. 26-28. The generalization of the equivalence of KSs for any case is straightforward using RD forms. KSs with the same $R_{x,y}$s and the same WT-PDFs for the connections in the RD form cannot be distinguished. Indistinguishable KSs with $R_{x,y}=2$ ( $x,y=on,off$ ) and tri-exponential $\phi_{on}(t)$ and $\phi_{off}(t)$ and the corresponding RD form are shown on Fig. 2E-2G.

Clearly, two KSs with different $R_{x,y}$s can be resolved by the analysis of a two-state trajectory. Among the advantageous of RD forms is by providing a powerful tool in resolving KSs with the same $R_{x,y}$s, and the same number of exponentials in $\phi_{on}(t)$ and $\phi_{off}(t)$, even without the need of performing actual calculations, based only on distinct complexity of the WT-PDFs for the connections in the corresponding RD forms, e.g. compare the KSs in Fig. 3A and Fig. 3B, or on different connectivity of RD forms, e.g. compare KSs in Figs. 3A-3B with the KS in Fig. 3C.

Perhaps it is worthwhile stressing that the above general statement implies that it is impossible to find positive (> 0) transition rates for the KSs in Figs. 3A-3C that make the $\phi_{x,y}(t_1,t_2)$s from these KSs the same, so these KSs can be distinguished by analyzing a two-state trajectory (excluding symmetric cases for which the $\phi_{x,y}(t_1,t_2)$s factorizes to the product of $\phi_x(t_1)\phi_y(t_2)$s).

Note that a RD form can preserve microscopic reversibility on the *on-off* level even when having irreversible connections. These can be balanced by the existence of direction dependent WT-PDFs for the connections. (Microscopic reversibility in a RD form means that the $\phi_{x,y}(t_1,t_2)$s obtained when reading the two-state trajectory in the forward direction are the same as the corresponding $\phi_{x,y}(t_1,t_2)$s obtained when reading the trajectory backwards, as suggested in Ref. 36 for aggregated Markov chains. Using matrix notation, microscopic reversibility means, $\phi_{x,y}(t_1,t_2) = [\phi_{y,x}(t_1,t_2)]^T$, where $T$ stands for the transpose of a matrix).

The division of KSs into equivalence groups (RD forms) is useful also when, on top of the information extracted from the 'original' two-state trajectory, additional information about the observed process is available. [Additional information can be inferred, under some physical assumptions, by analyzing different kind of measurements, e.g. the crystal structure of the biopolymer, or by analyzing two-state trajectories while varying some parameters, e.g. the substrate concentration (13-15)]. Suppose that the connectivity of the underlying KS is unchanged by the manipulation. Then, the additional information can be used to resolve KSs that correspond to the RD form found from the statistical analysis of the 'original' two-state trajectory, whereas any KS with a different RD form is irrelevant. Alternatively, when manipulating the system leads to a change in the connectivity of the underlying KS, or even to the addition or removal of substates, the RD forms obtained from the different data sets are distinct. Either of these possibilities is identifiable using RD forms and the corresponding KSs; in the first case an adequate parameter tuning relates the RD forms obtained from the various sources, whereas in the second case the RD forms cannot be related by a parameter tuning.

***Additional relationships between the data, the RD form, and the KS*** Additional relationships between the data, the RD form, and the underlying KS are discussed below when considering two cases. (*a*) All the $R_{x,y}$s are the same. For such cases, $R_{x,y}$ is the number of substates in each of the states in the RD form. Also, the number of exponents in $\phi_x(t)$ is the number of substates in state *x* in the (simplest) underlying KS. (*b*) Some of the $R_{x,y}$s are different. For such cases, the KS must have irreversible *on-off* connections and/or symmetry. (*b.1*) When $R_{on,off} \neq R_{off,on}$, there are irreversible *on-off* connections in the underlying KS. (*b.2*) When $R_{x,y} \geq R_{z,z}$ ($x \neq y$) for both values of *z=on, off*, $R_{x,y}$ is the number of substates in state *y* in the RD form. (*b.3*) When $R_{z,z} > R_{x,y}$ for the other three combinations of *x* and *y*, $R_{z,z}$ is the number of substates of both states in the RD form, and there is symmetry in state *z'* ($\neq z$) in the underlying KS. Take for example KS 3C, with the *on* to *off* transition rates having the same value. Then $R_{on,off} = R_{off,on} = R_{on,on} = 1$, and $R_{off,off} = 2$, but the topology of the RD form is the same as 3E. (*b.4*) When $R_{x,z} > R_{z,z}$ ($x \neq z$), there are irreversible *on-off* connections and a special connectivity in state *x* in the KS. In particular, $R_{z,z}$ is the *minimal* number of substates in state *x* of the KS among which the random walk must visit in each event in that state. Figure 4 shows an example for such a case, with $H_x = 0$ and no direct connections between substates in $\{N_x\}$ and substates in $\{M_x\}$.

***Concluding remarks*** The main effort in this paper is to utilize the information content in an ideal (noiseless, infinitely long) two-state trajectory for an efficient elucidation of a unique mechanism that can generate it. Accordingly, the KS space is partitioned into canonical forms that are (usually) not Markovian, where a canonical form is determined by the ranks of the $\phi_{x,y}(t_1,t_2)$s, and the (usually non-exponential) WT-PDFs for the connections among substates of different states in the canonical form. The relationships between the (fitting-free) $R_{x,y}$s, the KS's *on-off* connectivity, and the RD form's topology are the basis for our results, where the mathematical support is enclosed in Eqs. (1)-(2).

As a final remark, note that, in principle, one can collect successive *x-y* events in a selective way, such that the decomposition of the obtained two-dimensional histogram has *one* nonzero eigenvalue (SI.E). The number of these rank-one *x-y* histograms equal to the corresponding $R_{x,y}$, and are the terms in a particular external sum in Eq.(1) or Eq. (2). Although as $R_{x,y}$ increases it becomes harder to obtain these rank-one *x-y* histograms, they supply more details on the WT-PDFs for the connections in the RD form than their sum, and therefore can be viewed as the upper bound on the information content in a two-state trajectory.

* To whom correspondence should be addressed. E-mail: silbey@mit.edu.

Abbreviations: KS – kinetic scheme; WT-PDF – waiting time probability density function; MIR - manifest interconductance rank; BKU - Bauer–Kienker uncoupled; RD – reduced dimensions.

Conflict of interest statement: No conflicts declared.

**References**

1. Moerner, W. E. & Orrit, M. (1999) *Science* **283**, 1670-1676.

2. Weiss, S. (1999) *Science* **283**, 1676-1683.

3. Neher, E. & Sakmanm, B. (1976) *Nature* **260**, 799-802.

4. Kasianowicz, J. J., Brandin, E., Branton, D., & Deamer, D. W. (1996) *Proc. Natl. Acad. Sci. USA* **93**, 13770–13773.

5. Kullman, L., Gurnev, P. A., Winterhalter, M., & Bezrukov, S. M. (2006) *Phys. Rev. Lett.* **96**, 038101-038104.

6. Schuler, B., Lipman, E. A. & Eaton, W. A. (2002) *Nature* **419**, 743-747.

7. Yang, H., Luo, G., Karnchanaphanurach, P., Louie, T., Rech, I., Cova, S., Xun, L., Xie, X. S. (2003) *Science* **302**, 262-266.

8. Min, W., Lou, G., Cherayil, B. J., Kou, S. C., & Xie, X. S. (2005) *Phys. Rev. Lett.* **94**, 198302.

9. Rhoades, E., Gussakovsky, E. & Haran, G. (2003) *Proc. Natl. Acad. Sci. USA* **100**, 3197-3202.

10. Zhuang, X., Kim, H., Pereira, M. J. B., Babcock, H. P., Walter, N. G., & Chu, S. (2002) *Science* **296**, 1473-1476.

11. Lu, H., Xun, L. & Xie, X. S. (1998) *Science* **282**, 1877-1882.

12. Edman, L., Földes-Papp, Z., Wennmalm, S. & Rigler, R. (1999) *Chem. Phys*. **247**, 11-22.

13. Velonia, K., Flomenbom, O., Loos, D., Masuo, S., Cotlet, M., Engelborghs, Y., Hofkens, J., Rowan, A. E., Klafter, J., Nolte, R. J. M., *et al*. (2005) *Angew. Chem. Int. Ed.* **44**, 560-564.

14. Flomenbom, O., Velonia, K., Loos, D., Masuo, S., Cotlet, M., Engelborghs, Y., Hofkens, J., Rowan, A. E., Nolte, R. J. M., Van der Auweraer, M., *et al*. (2005) *Proc. Natl. Acad. Sci. USA*. **102**, 2368-2372.

15. English, B. P., Min, W., van Oijen, A. M., Lee, K. T., Luo, G., Sun, H., Cherayil, B. J., Kou, S. C., & Xie., X. S. (2006) *Nat. chem. Biol.* **2**, 87-94.

16. Nie, S., Chiu, D. T. & Zare R. N. (1994) *Science* **266**, 1018-1021.

17. Shusterman, R., Alon, S., Gavrinyov, T., & Krichevsky, O. (2004) *Phys. Rev. Lett.* **92**, 048303-1-4.

18. Zumofen, G., Hohlbein, J. & Hübner, C. G. (2004) *Phys. Rev. Lett.* **93**, 260601-1-4.

19. Cohen, A. E. & Moerner, W. E. (2006) *Proc. Natl. Acad. Sci. USA*. **103**, 4362-4365.

20. Dickson, R. M., Cubitt, A. B., Tsien, R. Y., & Moerner, W. E. (1997) *Nature* **388**, 355-358.

21. Chung, I. & Bawendi M. G. (2004) *Phys. Rev. B.* **70**, 165304-1-5.

22. Barkai, E., Jung, Y. & Silbey, R. (2004) *Annu. Rev. Phys. Chem.* **55**, 457-507.

23. Tang, J. & Marcus, R. A. (2005) *J. Chem. Phys*. **123**, 204511-1-6.

24. Horn, R. & Lange, K. (1983) *Biophys. J.* **43**, 207-223.

25. Qin. F., Auerbach, A. & Sachs, F. (2000) *Biophys. J.* **79**, 1915-1927.

26. Flomenbom, O., Klafter, J. & Szabo, A. (2005) *Biophys. J.* **88**, 3780-3783.

27. Flomenbom, O. & Klafter, J. (2005) *Acta Phys. Pol B* **36**, 1527-1535.

28. Flomenbom, O. & Klafter, J. (2005) *J. Chem. Phys.* **123**, 064903-1-10.

29. Witkoskie, J. B. & Cao, J. (2004) *J. Chem. Phys.* **121**, 6361-6372.

30. Bruno, W. J., Yang, J. & Pearson, J. (2005) *Proc. Natl. Acad. Sci. USA.* **102**, 6326-6331.

31. Bauer, R. J., Bowman, B. F. & Kenyon, J. L. (1987) *Biophys. J.* **52**, 961 – 978.

32. Kienker, P. (1989) *Proc. R. Soc. London B.* **236**, 269-309.

33. Fredkin, D. R. & Rice, J. A. (1986) *J. Appl. Prob.* **23**, 208-214.

34. Colquhoun, D. & Hawkes A. G. (1982) *Philos. Trans. R. Soc. Lond.* B *Biol. Sci.* **300**, 1–59.

35. Cao, J. (2000) *Chem. Phys. Lett.* **327**, 38-44.

36. Song, L. & Magdeby, K. L. (1994) *Biophys. J.* **67**, 91-104.

37. Vlad, M. O., Moran, F., Schneider, F. W., & Ross, J. (2002) *Proc. Natl. Acad. Sci. USA.* **99**, 12548-12555.

38. Yang, S. & Cao, J. (2002) *J. Chem. Phys.* **117**, 10996-11009.

39. Agmon, N. (2000) *J. Phys. Chem. B* **104,** 7830-7834.

40. Qian, H. & Elson, E. L. (2002) *Biophys. Chem.* **101**, 565-576.

41. Kou, S. C., Cherayil, B. J., Min, W., English, B. P., & Xie, X. S. (2005) *J. Phys. Chem. B* **109**, 19068-19081.

42. Sung, J. Y. & Silbey, R. J. (2005) *Chem. Phys. Lett.* **415**, 10-14.

43. Granek, R. & Klafter, J. (2005) *Phys. Rev. Lett.* **95**, 098106-1-4.

44. Shaevitz, J. W., Block, S. M., & Schnitzer, M. J. (2005) *Biophys. J.* **89**, 2277-2285.

45. Barsegov, V. & Thirumalai, D. (2005) *Phys. Rev. Lett.* **95**, 168302-1-4.

46. Šanda, F., & Mukamel, S. (2006) *J. Chem. Phys.* **108**, 124103-1-15.

47. Allegrini, P., Aquino, G., Grigolini, P., Palatella, L., & Rosa, A. (2003) *Phys. Rev. E* **68**, 056123-1-11.

48. Kolomeisky, A. B. & Fisher, M. E. (2000) *J. Chem. Phys.* **113**, 10867-10877.

49. Goychuk, I. & Hänggi, P. (2004) *Phys. Rev. E* **70**, 051915-1-9.

50. Flomenbom, O. & Klafter, J. (2005) *Phys. Rev. Lett.* **95**, 098105-1-4.

51. Yeramian, E. & Claverie, P. (1987) *Nature* **326**, 169-174.

**Figure Legends**

**FIG 1** A two-state trajectory (**A**) and the KS (**B**) (SI.E gives the technical details for generating the trajectory **A** corresponding to the KS **B**). Here, we consider noiseless, infinitely long, trajectories with prefect time resolution, which idealizes experimental trajectories. The reversible connection KS **B** has $L_{on}=3$ (squared substates), $L_{off}=10$ (circled substates), $N_{on}=M_{on}=2$ and $N_{off}=M_{off}=5$.

**FIG 2** Indistinguishable KSs. KSs **A-C** have the simplest RD form (**D**) of one substate in each of the states. KSs **A-C** are equivalent when they have the same $\phi_{on}$ and $\phi_{off}$. Equivalent KSs **E-F** have $R_{x,y}=2$, $x, y =$ *on, off*, and tri-exponential $\phi_{on}$ and $\phi_{off}$. The corresponding RD form is shown in **G**.

**FIG 3** Distinguishable KSs with $R_{x,y}=2$, $x, y =$ *on, off* and bi-exponential $\phi_{on}(t)$ and $\phi_{off}(t)$. (We exclude symmetry in this example). KS **3C** is distinct from KSs **3A** and **3B**, because the corresponding RD forms, **3E** and **3D**, respectively, have different connectivity. KS **3A** and KS **3B** are also distinct, because the WT-PDFs for the connections in the RD form of KS **3A** are exponential, whereas those of KS **3B** are direction-dependent and bi-exponentials.

**FIG 4** A KS with different $R_{x,y}$s, $R_{on,on}=2$, $R_{off,off}=1$, $R_{on,off}=4$, and $R_{off,on}=2$. The values for $R_{on,on}$ and $R_{off,off}$ are apparent in the functional form of the WT-PDFs for the connections. The KS is divided into two parts, **A** - *on* state, and, **B** - *off* state, for a convenient illustration. The filled substates are the initial ones, and those with directional arrows are the final ones (a directional arrow represents connections to all the initial substates of the other state). The striped substates in state *y* are those that contribute to the rank $R_{x,x}$; these substates are the minimal number of substates among which the random walk must visit in each event in the state. The RD form is shown on **C**.

# FIGURE 1 Silbey

A
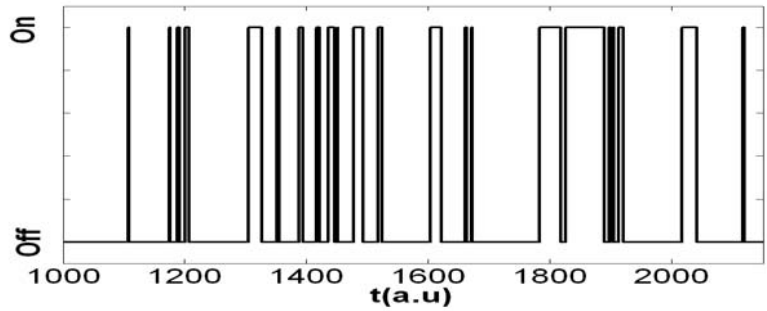

B
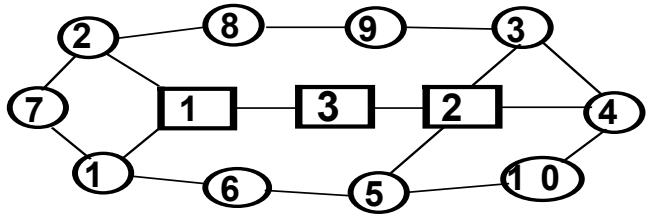

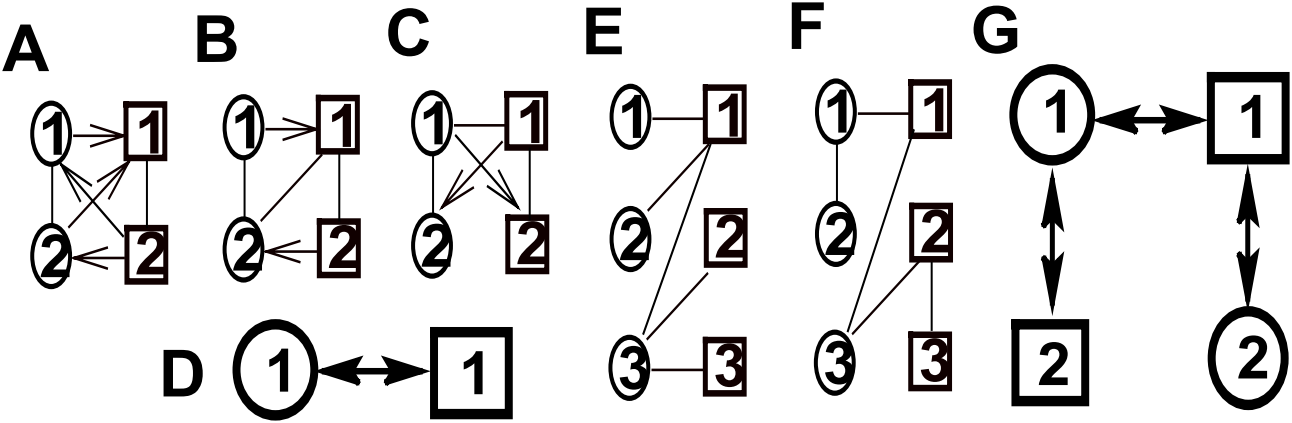

**FIGURE 2** Silbey

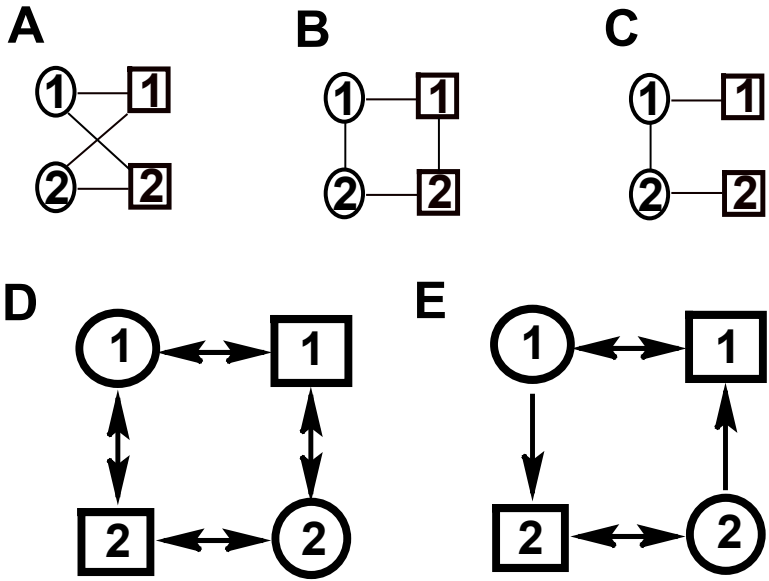

FIGURE 3 Silbey

# FIGURE 4 Silbey

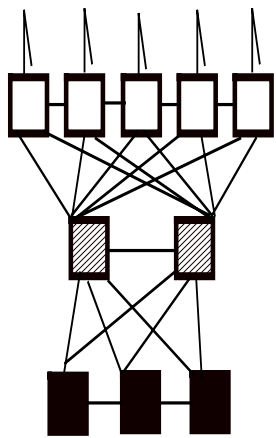 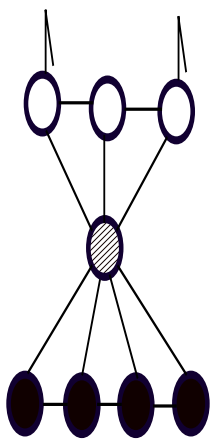 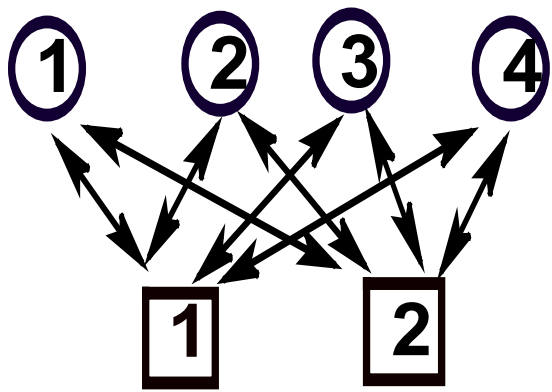

# Utilizing the information content in two-state trajectories

Ophir Flomenbom & Robert J. Silbey

## Supporting Information

### A

In this section, we give expressions for $\phi_x(t)$ and $\phi_{x,x}(t_1,t_2)$ using the formalism of the main text. Here, as in all the following sections, $x \neq y$, unless otherwise is explicitly indicated. The expression for $\phi_x(t)$ is obtained from Eq. (1) by integrating over one time argument, $\phi_x(t) = \int_0^\infty \phi_{x,y}(t,\tau)d\tau = \int_0^\infty \phi_{y,x}(\tau,t)d\tau$, which leads to,

$$\phi_x(t) = \sum_{n_y=1}^{N_y}\sum_{n_x=1}^{N_x} W_{n_x} f_{n_y n_x}(t) = \sum_{m_x \in \{M_x\}} \sum_{n_x=1}^{N_x}\sum_{n_y=1}^{N_y} W_{n_x} \tilde{f}_{m_x n_x}(t)\omega_{n_y m_x}. \qquad (A1)$$

The expression for $\phi_{x,x}(t_1,t_2)$ is obtained from Eq. (1) when introducing an additional summation that represents the random walk in state $y$ that takes place in between the two measured events in state $x$,

$$\phi_{x,x}(t_1,t_2) = \sum_{n'_x=1}^{N_x}\sum_{n_y=1}^{N_y}\sum_{n_x=1}^{N_x} W_{n_x} f_{n_y n_x}(t_1) p_{n'_x n_y} F_{n'_x}(t_2)$$

$$= \sum_{m_x \in \{M_x\}} \sum_{n'_x=1}^{N_x}\sum_{n_y=1}^{N_y}\sum_{n_x=1}^{N_x} W_{n_x} \tilde{f}_{m_x n_x}(t_1)\omega_{n_y m_x} p_{n'_x n_y} F_{n'_x}(t_2). \qquad (A2)$$

Here, $p_{n'_x n_y}$ is the probability that an event that starts at substate $n_y$ exits to substate $n'_x$, and is given by $p_{n'_x n_y} = \bar{f}_{n'_x n_y}(0)$, where $\bar{g}(s) = \int_0^\infty g(t)e^{-st}dt$ is the Laplace transform of $g(t)$.

Note that higher order successive WT-PDFs e.g. $\phi_{x,y,z}(t_1,t_2,t_3)$, do not contain additional information on top of the $\phi_{x,y}(t_1,t_2)$ s (1). When the underlying KS has no symmetry (i.e. the spectrum of $\phi_x(t)$, $x$ = on, off, is non-degenerate) and/or irreversible connections, it is sufficient to use $\phi_{x,y}(t_1,t_2)$ for $x \neq y$ (1), where for other cases, $\phi_{x,x}(t_1,t_2)$ s, $x$ = on, off, contain complementary information (2-4). For both cases, the rank-one two-dimensional PDFs that are contained in $\phi_{x,y}(t_1,t_2)$ constitute the upper bound on the information content in the two-state trajectory because their provide more details for constructing the RD form than their sum; See SI.E for further discussion regarding this point.

# B

In this section, we relate the formalism of the main text to that of the master equation. We start by writing $\phi_x(t)$ and $\phi_{x,y}(t_1,t_2)$ in matrix representation. $\phi_x(t)$ and $\phi_{x,y}(t_1,t_2)$ are given by,

$$\phi_x(t) = \vec{\mathbf{1}}_y^T \mathbf{V}_x \mathbf{G}_x(t) \mathbf{V}_y \vec{\mathbf{P}}_y(ss)/N_x, \tag{B1}$$

and,

$$\phi_{x,y}(t_1,t_2) = \vec{\mathbf{1}}_x^T \mathbf{V}_y \mathbf{G}_y(t_2) \mathbf{V}_x \mathbf{G}_x(t_1) \mathbf{V}_y \vec{\mathbf{P}}_y(ss)/N_x, \tag{B2}$$

where $N_x = \vec{\mathbf{1}}_x^T \mathbf{V}_y \vec{\mathbf{P}}_y(ss)$ and $\vec{\mathbf{1}}_x^T$ is the summation row vector of $1, L_x$ dimensions, $[\vec{\mathbf{1}}_x^T] = 1, L_x$. (The expression for $\phi_{x,x}(t_1,t_2)$ is obtained from Eq. (B1) when plugging in the factor $\mathbf{V}_y \overline{\mathbf{G}}_y(0)$, $\phi_{x,x}(t_1,t_2) = \vec{\mathbf{1}}_y^T \mathbf{V}_x \mathbf{G}_x(t_2) \mathbf{V}_y \overline{\mathbf{G}}_y(0) \mathbf{V}_x \mathbf{G}_x(t_1) \mathbf{V}_y \vec{\mathbf{P}}_y(ss)/N_x$). The quantities on the right hand side of Eqs. (B1) - (B2) are defined through the standard description of a random walk in an *on-off* KS (1-13). The time-dependent occupancy probabilities of state $x$ $\vec{\mathbf{P}}_x(t)$, $(\vec{\mathbf{P}}_x(t))_i = P_{x,i}(t)$, $i=1,...,L_x$, obey the reversible master equation:

$$\frac{\partial}{\partial t}\begin{pmatrix}\vec{\mathbf{P}}_{on}(t)\\ \vec{\mathbf{P}}_{off}(t)\end{pmatrix} = \begin{pmatrix}\mathbf{K}_{on} & \mathbf{V}_{off}\\ \mathbf{V}_{on} & \mathbf{K}_{off}\end{pmatrix}\begin{pmatrix}\vec{\mathbf{P}}_{on}(t)\\ \vec{\mathbf{P}}_{off}(t)\end{pmatrix}. \tag{B3}$$

In Eq. (B3), matrix $\mathbf{K}_x$, with dimensions $[\mathbf{K}_x] = L_x, L_x$, contains transition rates among substates in state *x*, and 'irreversible' transition rates from substates in state *x* to substates in state *y*. (The 'irreversible' transition rates are given on the diagonal, and are called irreversible because matrix $\mathbf{K}_x$ does not contain the back transition rates from state *y* to state *x*). Matrix $\mathbf{V}_x$, with dimensions $[\mathbf{V}_x] = L_y, L_x$, contains transition rates between states $x \to y$, where $(\mathbf{V}_x)_{ji}$ is the transition rate between substates $i_x \to j_y$. $\vec{\mathbf{P}}_x(ss)$ $[= \lim t \to \infty (\vec{\mathbf{P}}_x(t))]$ is the vector of occupancy probabilities in state *x* at steady state. (We assume that such a steady state exists, i.e. $L_{on} + L_{off} < \infty$). $\vec{\mathbf{P}}_x(ss)$ is found from Eq. (B3) for vanishing time derivative. $\mathbf{G}_x(t)$ in Eqs. (B1) - (B2) is the Green's function of state *x* for the irreversible process $\partial \mathbf{G}_x(t)/\partial t = \mathbf{K}_x \mathbf{G}_x(t)$, with the solution,

$$\mathbf{G}_x(t) = \exp[\mathbf{K}_x t] = \mathbf{X}\exp[\boldsymbol{\lambda}_x t]\mathbf{X}^{-1}. \tag{B4}$$

The second equality in Eq. (B4) follows from a similarity transformation $\boldsymbol{\lambda}_x = \mathbf{X}^{-1}\mathbf{K}_x\mathbf{X}$, and all the matrices in Eq. (B4) have dimensions $L_x, L_x$. No symmetry in state $x$ means non-degenerate eigenvalues in matrix $\boldsymbol{\lambda}_x$.

All the factors in Eq. (1), and Eqs. (A1)-(A2), can be expressed in terms of the matrixes of Eq. (B1). $W_{n_x}$ and $f_{n_y n_x}(t)$ are related to the master equation by,

$$W_{n_x} = \left(\mathbf{V}_y \vec{\mathbf{P}}_y(ss)\right)_{n_x} / N_x, \tag{B5}$$

and,

$$f_{n_y n_x}(t) = \left(\mathbf{V}_x \mathbf{G}_x(t)\right)_{n_y n_x}. \tag{B6}$$

$f_{n_y n_x}(t)$ can be further rewritten as,

$$f_{n_y n_x}(t) = \sum_{m_x} \omega_{n_y m_x} \widetilde{f}_{m_x n_x}(t), \tag{B7}$$

and similarly for $\left(\mathbf{V}_x \mathbf{G}_x(t)\right)_{n_y n_x}$ we have,

$$\left(\mathbf{V}_x \mathbf{G}_x(t)\right)_{n_y n_x} = \sum_k \left(\mathbf{V}_x\right)_{n_y k} \left(\mathbf{G}_x(t)\right)_{k n_x}. \tag{B8}$$

Note however that the factors in the sums Eq. (B7) and Eq. (B8) are not equal but proportional,

$$\widetilde{f}_{k n_x}(t) = \alpha_{x,k} \left(\mathbf{G}_x(t)\right)_{k n_x},$$

and

$$\omega_{n_y k} = \left(\mathbf{V}_x\right)_{n_y k} / \alpha_{x,k},$$

where

$$\alpha_{x,k} = -\left(\mathbf{K}_x\right)_{kk}.$$

## C

In this section, we give expressions for the WT-PDFs for the connections in the RD form, denoted by $\varphi_{j_y i_x}(t)$ and $\varphi_{i_x j_y}(t)$, for any KS. We do not consider symmetric KSs separately, but assume that the symmetry is apparent in the functional form of the $\varphi_{*^!*}(t)$ s. Further discussion regarding the topological interpretation of $\tilde{M}_x$ and $\tilde{N}_y$ used in Eq. (2) is also given.

The waiting time PDFs for the connections in the RD form are uniquely determined by the clustering procedure, specified in the section, *Mapping a KS into a RD form* (main text). The clustering procedure is based upon the identification of substates, in the interfaces of the *on-off* connectivity, that contribute to the ranks $R_{x,y}$. This makes the RD forms legitimate canonical forms that preserve all the information contained in the two-state trajectory.

The technical details to obtain the $\varphi_{*^!*}(t)$ s, given a KS, are spelled out below when considering two cases: (*1*) None of the terms in an external sum in Eq. (1), after the first or second equality, are proportional to each other, and (*2*) Some of the terms in an external sum in Eq. (1), after the first or second equality, are proportional to each other.

(*1.1*) *Reversible on-off connection KSs* Let $M_x \geq N_y$, or equivalently $N_x \geq M_y$. (Fig. D1A with *x=off*). Based on the clustering procedure, there are $N_y$ substates in each of the states in the RD form, and as many as $2N_y^2$ WT-PDFs for the connections in the RD form. Initial substates in state *x* are clustered, and the expression for $\varphi_{n_y i_x}(t)$ reads,

$$\varphi_{n_y i_x}(t) = \frac{1}{N_{x,m_y}} \sum_{n_x} P_{y,m_y}(ss)(\mathbf{V_y})_{n_x m_y} f_{n_y n_x}(t). \tag{C1}$$

In Eq. (C1), we use the normalization $N_{x,m_y}$, defined through the equations,

$$N_x = \vec{\mathbf{1}}_x^T \mathbf{V}_y \vec{\mathbf{P}}_y(ss) = \sum_{m_y, n_x} P_{y,m_y}(ss)(\mathbf{V_y})_{n_x m_y} = \sum_{m_y} N_{x,m_y} = \sum_{n_x} N_{x,n_x}.$$

As notation is concerned, we set in Eq. (C1) $j_y \to n_y$ because there are $n_y = 1,...,N_y$ substates in state *y* in the RD form, and we can also employ the meaning of $n_y$ as the initial substates in state *y* in the underlying KS. Additionally, we associate $m_y$ on the right hand side (RHS), which has the meaning of final substates in the underlying KS, with $i_x$ on the left hand side (LHS), i.e. $m_y \to i_x$. Note that for a KS with only reversible *on-off* connections, $m_y = 1,...,M_y$, so the values of $m_y$ and $i_x$ can be the same.

The expression for $\varphi_{i_x n_y}(t)$ is different than that for $\varphi_{n_y i_x}(t)$ in both the normalization used and the factors that are summed, which is a result of the mapping of the initial substates in state $y$ to themselves. $\varphi_{i_x n_y}(t)$ is given by,

$$\varphi_{i_x n_y}(t) = \frac{1}{N_{y,n_y}} \sum_{m_x} P_{x,m_x}(ss)(\mathbf{V_x})_{n_y m_x} \tilde{f}_{m_y n_y}(t)\tilde{\omega}_{m_y} \quad ; \quad \tilde{\omega}_{m_y} = \sum_{n_x} \omega_{n_x m_y} . \quad (C2)$$

Note that here, $\varphi_{i_x n_y}(t) = \tilde{f}_{m_y n_y}(t)\tilde{\omega}_{m_y} = (\mathbf{G_y}(t))_{m_y n_y} \sum_{n_x} (\mathbf{V_y})_{n_x m_y}$. In Eq. (C2), we associate $m_y$ on the RHS with $i_x$ on the LHS, i.e. $m_y \rightarrow i_x$. Again, for a KS with only reversible transitions, the $i_x$s can have the same values as of the $m_y$s.

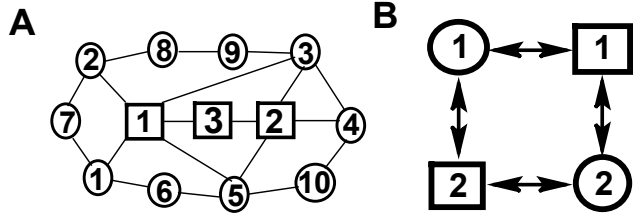

**FIG C1 A** – A reversible connection KS, with $N_{on}= M_{on}=2$ and $N_{off}=M_{off}=5$. **B** - The RD form of KS **A**. The RD form's substate $1_{off}$ corresponds to the cluster of the KS's *off* substates $1_{off}$-$3_{off}$ and $5_{off}$, because these are connected to substate $1_{on}$ in the KS. The RD form's substate $2_{off}$ corresponds to the cluster of the KS's *off* substates $3_{off}$-$5_{off}$, because these are connected to substate $2_{on}$ in the KS. Note that a particular initial substate can appear in more than a single cluster, which simply means that the overall steady-state flux into the substate in divided into several contributions. The initial *on* substates in the KS both contribute to $R_{off,on}$ so they are mapped to themselves in the RD form. The WT-PDFs for the connections can be obtained from Eqs. (C1)-(C2).

(*1.2*) *Irreversible on-off connection KSs* There are no conceptual differences in obtaining the $\varphi_{*,*}(t)$s for irreversible versus reversible *on-off* connection KSs. The reason is that the clustering procedure is based on the directional connections between final substates in state $x$ and initial substates in state $y$. However, some technical details may differ. We consider two cases.

(*1.2.1*) Let $M_x \geq N_y$ and $M_y \geq N_x$. (Fig. C2A-C2B). Then, the WT-PDFs for the connections are given by,

$$\varphi_{n_y n_x}(t) = \frac{1}{N_{x,n_x}} \sum_{m_y} P_{y,m_y}(ss)(\mathbf{V_y})_{n_x m_y} f_{n_y n_x}(t) = f_{n_y n_x}(t), \tag{C3}$$

and,

$$\varphi_{n_x n_y}(t) = \frac{1}{N_{y,n_y}} \sum_{m_x} P_{x,m_x}(ss)(\mathbf{V_x})_{n_y m_x} f_{n_x n_y}(t) = f_{n_x n_y}(t). \tag{C4}$$

Note that for this case any $\varphi_{**}(t)$ equal to the corresponding $f_{**}(t)$. This is an outcome of the KS's topology for which in both the *on* to *off* and the *off* to *on* connections, the number of initial substates in a given state is lower than the number of final substates in the other state.

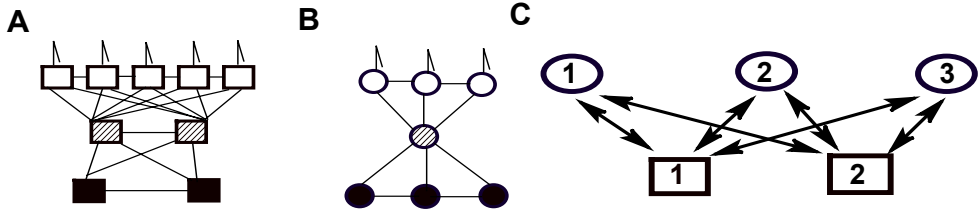

**FIG C2** An example for a KS with irreversible *on-off* connections, and $N_{on}=2$, $M_{on}=5$, $N_{off}=3$, and $M_{off}=3$. The KS is divided into two parts shown on **A** (*on* state) and **B** (*off* state) for a convenient illustration. The RD form is shown on **C**. The WT-PDFs for the connections can be obtained from Eqs. (C3) - (C4).

*(1.2.2)* Let $N_x > M_y$ and $N_y > M_x$. (Fig. C3A-C3B). Then, the WT-PDFs for the connections are given by,

$$\varphi_{j_y i_x}(t) = \frac{1}{N_{x,m_y}} \sum_{n_x} P_{y,m_y}(ss)(\mathbf{V_y})_{n_x m_y} \tilde{f}_{m_x n_x}(t) \tilde{\omega}_{m_x}, \tag{C5}$$

and,

$$\varphi_{i_x j_y}(t) = \frac{1}{N_{y,m_x}} \sum_{n_y} P_{x,m_x}(ss)(\mathbf{V_x})_{n_y m_x} \tilde{f}_{m_y n_y}(t) \tilde{\omega}_{m_y}. \tag{C6}$$

In Eqs. (C5)-(C6), we use the mapping $m_y \to i_x$ and $m_x \to j_y$ between the RHS and the LHS indexes. (In particular, $m_y - (N_y - H_y) = i_x$, and $m_x - (N_x - H_x) = j_y$).

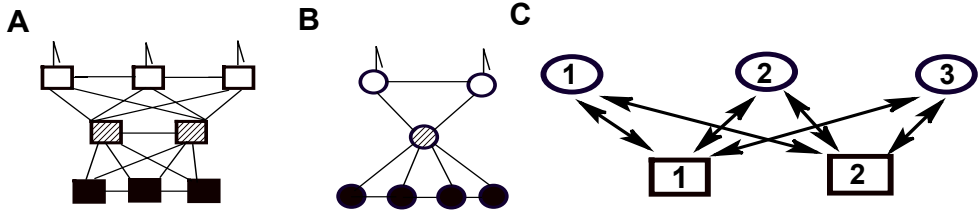

**FIG C3** An irreversible *on-off* connection KS with $N_{on}=3$, $M_{on}=3$, $N_{off}=4$, and $M_{off}=2$. The panels are divided as in Fig. C2. The WT-PDFs for the connections can be obtained from Eqs. (C5) - (C6).

(*2*) We turn now to treat cases for which some of the terms in Eq. (1) are proportional. We consider only KSs with reversible *on-off* connections, but the same ideas can be applied for KSs with irreversible *on-off* connections.

Let $M_x \leq N_y$, or equivalently $N_x \leq M_y$. (See Fig. C4 with $x=off$). So it follows that, $R_{x,y} < M_x$, which is a result of a special *on-off* connectivity. In particular, let $\{O_y\}$ and $\{O_x\}$ be the groups of substates in states $y$ and $x$ respectively, such that the substates in $\{O_x\}$ are connected only to the substates in $\{O_y\}$, and $O_y < O_x$. (In Fig. C4A, the group $\{O_{off}\}$ contains the substates $1_{off}$, $2_{off}$ and $3_{off}$, and the group $\{O_{on}\}$ contains the substates $1_{on}$ and $2_{on}$). Thus, both initial and final substates contribute to the rank $R_{z,z'}$ for $z \neq z'$, and the expressions for the $\varphi_{*,*}(t)$s are distinct in each of the following three regimes:

(*a*) For $n_x \notin \{O_x\}$ and $n_y \notin \{O_y\}$,

$$\varphi_{j_y i_x}(t) = \frac{1}{N_{x,n_x}} \sum_{m_y} P_{y,m_y}(ss)(\mathbf{V_y})_{n_x m_y} \tilde{f}_{m_x n_x}(t) \sum_{n_y \notin \{O_y\}} \omega_{n_y m_x}, \qquad (C8)$$

and,

$$\varphi_{i_x j_y}(t) = \frac{1}{N_{y \in O_y, m_x}} \sum_{n_y} P_{x,m_x}(ss)(\mathbf{V_x})_{n_y m_x} f_{n_x n_y}(t), \qquad (C9)$$

where $N_{y \in O_y, m_x} = \sum_{n_y \in \{O_y\}} P_{x,m_x}(ss)(\mathbf{V_x})_{n_y m_x}$, and we associate $n_x \to i_x$ and $m_x \to j_y$.

(*b*) For $n_x \notin \{O_x\}$ and $n_y \in \{O_y\}$,

$$\varphi_{j_y i_x}(t) = \frac{1}{N_{x,n_x}} \sum_{m_y} P_{y,m_y}(ss)(\mathbf{V_y})_{n_x m_y} f_{n_y n_x}(t), \qquad (C10)$$

and,

$$\varphi_{i_x j_y}(t) = \frac{1}{N_{y,n_y}} \sum_{m_x} P_{x,m_x}(ss)(\mathbf{V}_x)_{n_y m_x} f_{n_x n_y}(t), \quad (C11)$$

where we associate $n_y \to j_y$ and $n_x \to i_x$.

(c) For $n_x \in \{O_x\}$ and $n_y \in \{O_y\}$,

$$\varphi_{i_x j_y}(t) = \frac{1}{N_{y,n_y}} \sum_{m_x} P_{x,m_x}(ss)(\mathbf{V}_x)_{n_y m_x} \tilde{f}_{m_y n_y}(t) \sum_{n_x \in \{O_x\}} \omega_{n_x m_y}, \quad (C12)$$

and,

$$\varphi_{j_y i_x}(t) = \frac{1}{N_{x \in O_x, m_y}} \sum_{n_x \in \{O_x\}} P_{y,m_y}(ss)(\mathbf{V}_y)_{n_x m_y} f_{n_y n_x}(t), \quad (C13)$$

where we associate $n_y \to j_y$ and $m_y \to i_x$.

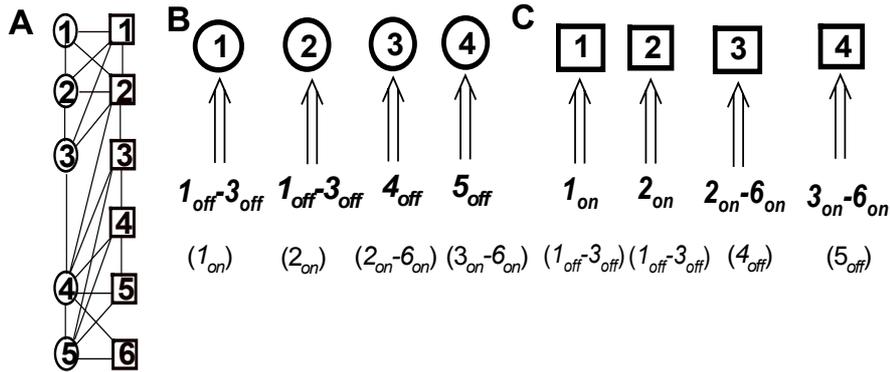

**FIG C4 A** A reversible connection KS with $R_{x,y}=4$ ($x \neq y$). The RD form's topology is shown on **B-C**. The clustering procedure and the parent substates (in the parenthesis) are indicated at the base of the double arrows. For example, substate $1_{off}$ in the RD form corresponds to the cluster of initial-*off*-substates $1_{off}$-$3_{off}$ in the KS. These are connected to substate $1_{on}$ in the KS. The WT-PDFs for the connections in the RD form can be obtained from Eqs. (C8)-(C13).

Now, we use $O_y$ and $O_x$ for expressing $R_{x,y}$. When $M_x < N_y$ and $\{O_x\}$ and $\{O_y\}$ are as defined above,

$$R_{x,y} = M_x - (O_x - O_y). \quad (C14)$$

This result can be generalized to the case of $J$ groups in the underlying KS that are connected in the way defined above for the case of a single pair of groups. The generalized result reads,

$$R_{x,y} = M_x - \sum_j (O_{x,j} - O_{y,j}). \tag{C15}$$

These expressions imply that $\widetilde{M}_x$ and $\widetilde{N}_y$ on Eq. (2) are related to the KS's topology by,

$$\widetilde{M}_x = M_x - \sum_j O_{x,j}, \tag{C16}$$

and,

$$\widetilde{N}_y = \sum_j O_{y,j}. \tag{C17}$$

When $M_x > N_y$, and there are groups $\{Z_x\}$ and $\{Z_y\}$, with $Z_x < Z_y$, such that substates in $\{Z_y\}$ are connected only to substates in $\{Z_x\}$, we define $O_x = M_x - Z_x$ and $O_y = N_y - Z_y$, and Eq. (C14) holds. For $J$ such groups, we define $O_{x,j} = M_x/J - Z_{x,j}$ and $O_{y,j} = N_y/J - Z_{y,j}$, and Eqs. (C15)-(C17) hold.

For a KS with symmetry, $\widetilde{M}_x$ and $\widetilde{N}_y$ are chosen in a different way than the one relies on the *on-off* connectivity; for such a case, the choice that makes the number of additives in the external sums of Eq. (2) minimal simply groups the identical PDFs. The topology of the RD form, however, is determined by the larger $R_{x,y}$; See the section *Additional relationships between the data, the RD form, and the KS* in main text for further discussion.

## **D**

In this section, a discussion about the rank of an underlying KS is given, using the master equation formalism. This will be shown to emphasize the advantageous of the formalism of the main text. Following Fredkin and Rice (1), we introduce the singular value decomposition of $\mathbf{V}_x$,

$$\mathbf{V}_x = \mathbf{u}_x \mathbf{S}_x \mathbf{v}_x^T, \tag{D1}$$

where $[\mathbf{u}_x] = L_y, L_y$, $[\mathbf{S}_x] = L_y, L_x$, and $[\mathbf{v}_x^T] = L_x, L_x$. Let $r_{x \to y}$ ($x \neq y$) be the number of non-zero singular values of $\mathbf{S}_x$. ($r_{x \to y}$ is the rank used by Pearson and collaborators in Ref. 8). Using Eq. (D1), we rewrite Eq. (B2) for $x=on$,

$$\phi_{on,off}(t_1,t_2) = \left(\vec{\mathbf{1}}_{on}^T \mathbf{u}_{off}\right) \mathbf{S}_{off}\left(\mathbf{v}_{off}^T \mathbf{G}_{off}(t_2) \mathbf{u}_{on}\right) \mathbf{S}_{on}\left(\mathbf{v}_{on}^T \mathbf{G}_{on}(t_1) \mathbf{u}_{off}\right) \mathbf{S}_{off}\left(\mathbf{v}_{off}^T \vec{\mathbf{P}}_{off}(ss)/N_{off}\right)$$

$$\equiv \left(\vec{\mathbf{A}}^T\right)_{1,L_{on}} \mathbf{S}_{off}\left(\mathbf{B}(t_2)\right)_{L_{off},L_{off}} \mathbf{S}_{on}\left(\mathbf{C}(t_1)\right)_{L_{on},L_{on}} \mathbf{S}_{off}\left(\vec{\mathbf{D}}\right)_{L_{off},1}$$

$$= \left(\vec{\mathbf{A}}^T\right)_{1,L_{on}} \left(\mathbf{S}_{off}\left(\mathbf{B}(t_2)\right)_{L_{off},L_{off}}\right)\left(\mathbf{S}_{on}\left(\mathbf{C}(t_1)\right)_{L_{on},L_{on}}\right)\left(\mathbf{S}_{off}\left(\vec{\mathbf{D}}\right)_{L_{off}*1}\right). \tag{D2}$$

The last three factors on the third line of Eq. (D2) are,

$$\mathbf{S}_{off}\left(\mathbf{B}(t_2)\right)_{L_{off},L_{off}} = \begin{pmatrix} \beta_{11}(t_2) & : & : & : & \beta_{1L_{off}}(t_2) \\ : & : & : & : & : \\ \beta_{r_{off \to on}1}(t_2) & : & : & : & \beta_{r_{off \to on}L_{off}}(t_2) \\ 0 & 0 & 0 & 0 & 0 \\ 0 & 0 & 0 & 0 & 0 \end{pmatrix}_{L_{on},L_{off}} \tag{D3.1}$$

$$\mathbf{S}_{on}\left(\mathbf{C}(t_1)\right)_{L_{on},L_{on}} = \begin{pmatrix} \gamma_{11}(t_1) & : & : & : & \gamma_{1L_{on}}(t_1) \\ : & : & : & : & : \\ \gamma_{r_{on \to off}1}(t_1) & : & : & : & \gamma_{r_{on \to off}L_{on}}(t_1) \\ 0 & 0 & 0 & 0 & 0 \\ 0 & 0 & 0 & 0 & 0 \end{pmatrix}_{L_{off},L_{on}} \tag{D3.2}$$

$$\mathbf{S}_{off}\left(\vec{\mathbf{D}}\right)_{L_{off},1} = \begin{pmatrix} \delta_1 \\ : \\ \delta_{r_{off \to on}} \\ 0 \\ 0 \end{pmatrix}_{L_{on},1} \tag{D3.3}$$

Using Eqs. (D3.1)-(D3.3) in Eq. (D2) leads to,

$$\phi_{on,off}(t_1,t_2) = \begin{pmatrix} \alpha_1(t_2) & : & : & \alpha_{L_{off}}(t_2) \end{pmatrix}_{1,L_{off}} \begin{pmatrix} \varepsilon_1(t_1) \\ : \\ \varepsilon_{r_{on \to off}}(t_1) \\ 0 \\ 0 \end{pmatrix}_{L_{off},L_{on}} = \sum_{i=1}^{r_{on \to off}} \alpha_i(t_2)\varepsilon_i(t_1). \tag{D4}$$

Now, dealing with data means that the time is discrete, namely, $\phi_{on,off}(t_1,t_2)$ is a matrix, and $r_{on \to off}$ in Eq. (D4) may be interpreted as the rank of the matrix $\phi_{on,off}(t_1,t_2)$, i.e. the number of non-zero eigenvalues (or singular values for a non-square matrix) of matrix $\phi_{on,off}(t_1,t_2)$ is $r_{on \to off}$. Although Eq. (D4) is also valid for a KS with a different number of non-zero singular values of $\mathbf{V}_{on}$ and $\mathbf{V}_{off}$, $r_{on \to off} \neq r_{off \to on}$, namely for a KS with (*a*) irreversible *on-off* connections, and (*b*) symmetry, the above interpretation of $r_{on \to off}$ doesn't always hold for cases (*a*)-(*b*). Take for example,

$\alpha_i(t_2)\varepsilon_i(t_1) = \alpha(t_2)\varepsilon_i(t_1)$,

which leads to

$\phi_{on,off}(t_1,t_2) = \alpha(t_2)\sum_i \varepsilon_i(t_1)$.

Namely, the number of non-zero eigenvalues, or singular values, that are obtained from the experimental $\phi_{on,off}(t_1,t_2)$ is one, $R_{on,off}=1$, but $r_{on \to off}$ may be larger than one.

    Although Eq. (1), Eq. (B2) and Eq. (D4) are mathematically equivalent, Eq. (1), which highlights the importance of the *on-off* connectivity in expressing the two-dimensional histograms, is more useful than the other two representations of $\phi_{x,y}(t_1,t_2)$ in relating the data to the topology and the *on-off* connectivity of the KS. Firstly, Eq. (1) emphasizes the unique role of the initial and final substates in each of the states. Then, it gives explicit meaning to the factors that are contained in $\phi_{x,y}(t_1,t_2)$. For example, the above scenario corresponds to the case $F_{n_y}(t_2) = F(t_2)$, which has much clearer physical meaning than $\alpha_i(t_2) = \alpha(t_2)$. ($F_{n_y}(t)$ is the WT-PDF, conditional on starting an event in substate $n_y$, for exiting to any initial substate of state *x*). Additionally, other cases of factorizations of $\phi_{x,y}(t_1,t_2)$ to the product of the individual event WT-PDFs,

$\phi_{x,y}(t_1,t_2) = \phi_x(t_1)\phi_y(t_2)$, for special topologies were found using Eq. (1) (2-4). As illustrated here, Eq. (1) forms the basis for relating any combination of $R_{x,y}$ s to the KS *on-off* connectivity and details.

**E**

In this section, we suggest a method to perform a simulation of a random walk in a RD form, and discuss ways to build the RD form from the data. Complementary to the general results given in the main text, a particular emphasis is put on the technical details for restoring the RD form from data, where also given is a treatment for finite trajectories.

To generate a two-state trajectory from a random walk in a RD form, a modified Gillespie Mote-Carlo method can be used. Each transition in the simulation happens in two steps. Assume the process starts at substate $i_x$. The first step chooses the destination of the next location, determined by the weights of making a transition $i_x \rightarrow j_y$: $w_{j_y i_x} = \overline{\varphi}_{j_y i_x}(0) / \sum_{j'_y} \overline{\varphi}_{j'_y i_x}(0)$. (Note that from the analytical expressions of section **C**, the sum $\sum_{j'_y} \overline{\varphi}_{j'_y i_x}(0)$ is unity, but due to numerical issues it can be smaller than unity). The second step uses the particular $j_y$, and draws a random time out of a normalized density $\varphi_{j_y i_x}(t) / \overline{\varphi}_{j_y i_x}(0)$. The procedure is then repeated at the new location. Note that, in principle, it is faster to generate a two-state trajectory using the RD form rather than the underlying KS.

Methods for restoring the RD form's topology and $\varphi_{*,*}(t)$ s from the data are suggested. Firstly, recall that the algorithm for finding the RD form from the data reads: (*a*) Find the number of substates in the RD form using decomposition of the $\phi_{x,y}(t_1,t_2)$ s. (*b*) Obtain the spectrum of the $\phi_x(t)$ s using fitting procedures. The spectrum of the $\varphi_{j_y i_x}(t)$ s is the same spectrum as that of $\phi_x(t)$, because substates of the same state are not connected in the RD form. Differences between the $\varphi_{*,*}(t)$ s and the $\phi_*(t)$ s lay in the pre-exponential coefficients. (Steps (*a*)-(*b*) can be permutated). (*c*) Apply a fitting procedure for finding the pre-exponential coefficients of the $\varphi_{*'*}(t)$ s.

Note that determining the RD form means extracting all the information in the data. Further conclusions on the underlying mechanism should be based on additional sources of information.

Figure 1A (main text) shows a two-state trajectory (with $dt = 0.5\, a.u.$) generated from the RD form shown in Fig. 3D (main text), corresponding to the underlying KS shown in Fig. 1B (main text). The above algorithm was used to generate the data. All the transition rates in the underlying KS are set to be the same, $\lambda = (50\, a.u.)^{-1}$, which leads to $<t_{on}> = 30\, a.u.$ and $<t_{off}> = 100\, a.u.$ (these numbers are rounded to within less than 2%). The WT-PDFs for the connections are found using Eqs. (C1) - (C2), and are shown in Fig. E1.

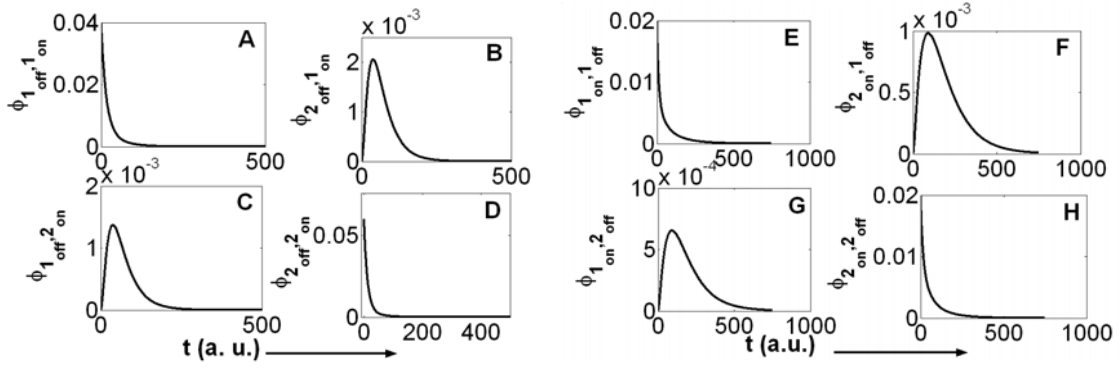

**FIG E1** The WT-PDFs for the connections in the RD form (Fig. 3E, main text) for *on* to *off* connections (**A-D**) and for *off* to *on* connections (**E-F**). The corresponding KS is shown in Fig. 1B (main text), where all the transition rates are set to be the same, $(50\, a.u.)^{-1}$. (The WT-PDFs are labeled with $\phi$ instead of $\varphi$ due to technical difficulties).

Figure E2 displays $\phi_{on}(t)$ and $\phi_{off}(t)$ constructed, on one hand, using Eq. (B1), and on the other hand, using a $10^6$ *on-off* events trajectory. Although the KS has fairly many substates, this number of *on-off* events is sufficiently large, so that the $\phi_x(t)$ s are accurately obtained for times such that their amplitudes are 2 orders of magnitudes smaller than its' maximal values.

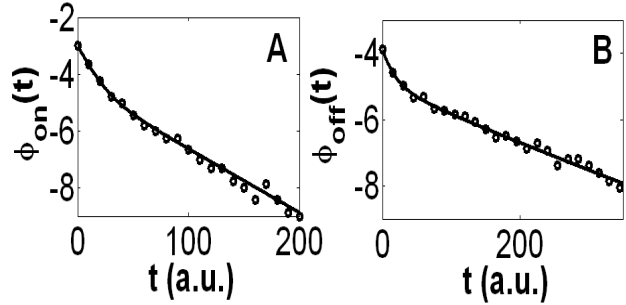

**FIG E2** The WT-PDFs of the *on* (**A**) and the *off* (**B**) states on a log-linear scale. Shown are both the WT-PDFs obtained from a numerical solution of Eq. (B1), full line, and by constructing these PDFs from a $10^6$ *on-off* event trajectory, circled symbols.

To get the $R_{x,y}$ s, we apply singular value decomposition (SVD) on the numerically obtained $\phi_{x,y}(t_1,t_2)$ s, and recover $R_{x,y} = 2$ ($\forall x, y = on, off$). The ratio of the large to small singular value is large (~250). Figure E3A shows $\phi_{on,off}(t_1,t_2)$ found from Eq. (B2). The two components of the SVD of $\phi_{on,off}(t_1,t_2)$ are shown on Fig. E3C (large singular value) and Fig. E3D (small singular value). The sum of the two components is shown in Fig. E3B, and recovers the original $\phi_{on,off}(t_1,t_2)$ (Fig. E3A). As implied by the large ratio of the two singular values, the contribution from the large singular value contains most of the signal (Fig. E3C). The two-dimensional function from the small singular value is not positive always (Fig. E3B), which immediately means that it is not a PDF.

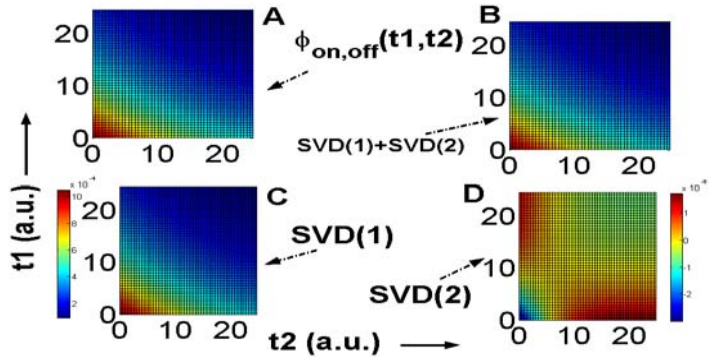

**FIG E3** The two dimensional histogram $\phi_{on,off}(t_1,t_2)$ (**A**), and the two contributions from the SVD (**C & D**), and their sum (**B**). Note that the function in **D** is not positive everywhere (see text for discussion).

The fact that the function from the second singular value in this example is not positive everywhere implies that the decomposition of a matrix $\phi_{x,y}(t_1,t_2)$ has, here, a limited use. Eigenvalue, or singular value, decompositions preserve the rank of the original matrix. Thus, the $R_{x,y}$ s are correctly obtained from such decompositions, and, as discussed in the main text, are related to the KS *on-off* connectivity and details. However, the rank-one two-dimensional histograms, which partition $\phi_{x,y}(t_1,t_2)$ according to Eq. (1) or Eq. (2), are not the rank-one two-dimensional functions of the decomposition. The correct division of the matrix $\phi_{x,y}(t_1,t_2)$ into the rank-one two-dimensional histograms can, in principle, be obtained from the data by collecting specific successive *on-off* events. For a 2 by 2 RD form, the rank-one two-dimensional PDFs that are contained in $\phi_{on,off}(t_1,t_2)$ are given by,

$$C1_{on,off}(t_1,t_2) = \left(W_{1_{on}}\varphi_{1_{off}1_{on}}(t_1) + W_{2_{on}}\varphi_{1_{off}2_{on}}(t_1)\right)\left(\varphi_{1_{on}1_{off}}(t_2) + \varphi_{2_{on}1_{off}}(t_2)\right), \quad (E1)$$

and,

$$C2_{on,off}(t_1,t_2) = \left(W_{1_{on}}\varphi_{2_{off}1_{on}}(t_1) + W_{2_{on}}\varphi_{2_{off}2_{on}}(t_1)\right)\left(\varphi_{1_{on}2_{off}}(t_2) + \varphi_{2_{on}2_{off}}(t_2)\right), \quad (E2)$$

with,

$$\phi_{on,off}(t_1,t_2) = C1_{on,off}(t_1,t_2) + C2_{on,off}(t_1,t_2). \quad (E3)$$

Similar equations can be written for the other $\phi_{x,y}(t_1,t_2)$ s. Figure E4 shows $C1_{on,off}(t_1,t_2)$ and $C2_{on,off}(t_1,t_2)$ for the specific example discussed above. The utility of these rank-one histograms is by supplying additional information for extracting the $\varphi_{**}(t)$ s (for example, the second factor in each of the Eqs. (E1)-(E2) is normalized to one, so integrating over $t_2$ gives exactly the first factor).

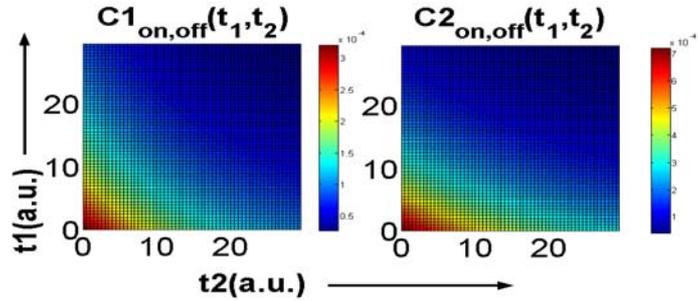

**FIG E4** The rank-one two dimensional PDFs that are contained in $\phi_{on,off}(t_1,t_2)$ (Fig. **E3A**). Eq. (E1)-(E2) were used to obtain these PDFs. Compare these PDFs with the functions in Figs. E3C-E3D.

The numerical result that shows that the contribution from the large singular value contains most of the signal corresponds to the limit of an infinitely long trajectory. Thus, one may expect technical difficulties in detecting the exact number of nonzero singular values from an experimental matrix, due to the limited number of events in the trajectory. Indeed, SVD is frequently used to filter noise by throwing away the components from the singular values that are smaller than a user-defined cut-off. Here, we wish to determine the number of nonzero singular values that would appear in the matrix obtained from an infinitely long trajectory by analyzing the matrix obtained from a finite trajectory, so any cut-off should be carefully defined. We suggest a simple manipulation that can be used to define a cut-off, but note that in ambiguous cases a Bayesian Information Criteria, according to which the optimum complexity of a model given data can be determined, should be applied for constructing the RD form's topology from the finite trajectory. The simplest way to smooth a PDF obtained from data is to plot it with a larger bin size. This doesn't change the rank of the original matrix, as long as the smaller *dimension* of the obtained matrix is not smaller than the *rank* of the original matrix.

Figure E5A shows $\phi_{on,off}(t_1,t_2)$ obtained from a two-state trajectory of $10^6$ *on-off* events, calculated for events in the range $0\ a.u. \leq t_1, t_2 \leq 25\ a.u.$, where the bin size equal $dt$. The first four rations of successive singular values of this matrix are, 35.3, 1.07, 1.01, 1.07. (The singular values are arranged in a descending order). This result is quite ambiguous. Figure E5B shows $\phi_{on,off}(t_1,t_2)$ with the bin size taken to be $50dt$, calculated for events in the range $0\ a.u. \leq t_1, t_2 \leq 100\ a.u.$ The first three ratios of successive singular values of this matrix are, 458, 7.9, 1.73. This may be interpreted as a rank-two matrix. (Similar results are obtained for the ratios between the sums of the absolute value of the rank-one functions of successive singular values).

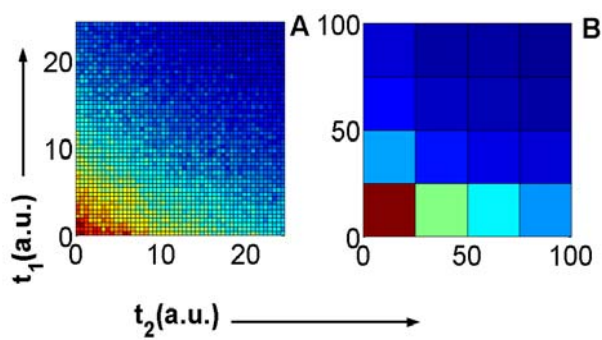

**FIG E5** Two-dimensional histograms obtained from a $10^6$ *on-off* event trajectory. **A** – $\phi_{on,off}(t_1,t_2)$ with a bin size equal $dt$, for events in the range $0\ a.u. \leq t_1, t_2 \leq 25\ a.u.$ **B** – $\phi_{on,off}(t_1,t_2)$ with a bin size of $50dt$, for events in the range, $0\ a.u. \leq t_1, t_2 \leq 100\ a.u.$ See text for discussion.

Lastly, the final step of the algorithm, after determining the RD form's topology and the spectrum of the $\phi_*(t)$s, is to apply maximum likelihood methods to find the coefficients of the $\varphi_{*'*}(t)$s. Self consistent tests include constructing the $\phi_*(t)$s and the $\phi_{*',*}(t_1,t_2)$, and the rank-one two-dimensional histograms if these are available, using the obtained $\varphi_{*'*}(t)$s.

**References**

1. Fredkin, D. R. & Rice, J. A. (1986) *J. Appl. Prob.* **23**, 208-214.

2. Flomenbom, O., Klafter, J. & Szabo, A. (2005) *Biophys. J.* **88**, 3780-3783.

3. Flomenbom, O. & Klafter, J. (2005) *Acta Phys. Pol B* **36**, 1527-1535.

4. Flomenbom, O. & Klafter, J. (2005) *J. Chem. Phys.* **123**, 064903-1-10.

5. Horn, R. & Lange, K. (1983) *Biophys. J.* **43**, 207-223.

6. Qin, F., Auerbach, A. & Sachs, F. (2000) *Biophys. J.* **79**, 1915-1927.

7. Witkoskie, J. B. & Cao, J. (2004) *J. Chem. Phys.* **121**, 6361-6372.

8. Bruno, W. J., Yang, J. & Pearson, J. (2005) *Proc. Natl. Acad. Sci. USA*. **102**, 6326-6331.

9. Bauer, R. J., Bowman, B. F. & Kenyon, J. L. (1987) *Biophys. J.* **52**, 961 – 978.

10. Kienker, P. (1989) *Proc. R. Soc. London B*. **236**, 269-309.

11. Colquhoun, D. & Hawkes A. G. (1982) *Philos. Trans. R. Soc. Lond.* B *Biol. Sci.* **300**, 1–59.

12. Cao, J. (2000) *Chem. Phys. Lett.* **327**, 38-44.

13. Song, L. & Magdeby, K. L. (1994) *Biophys. J.* **67**, 91-104.

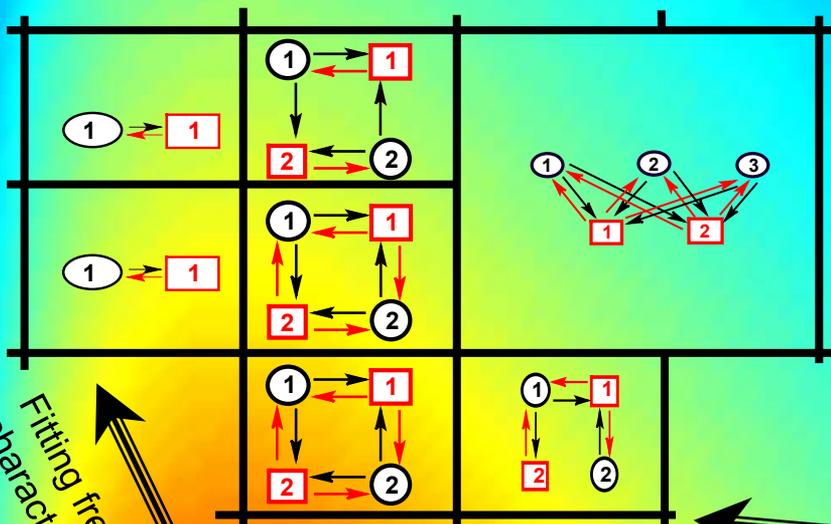
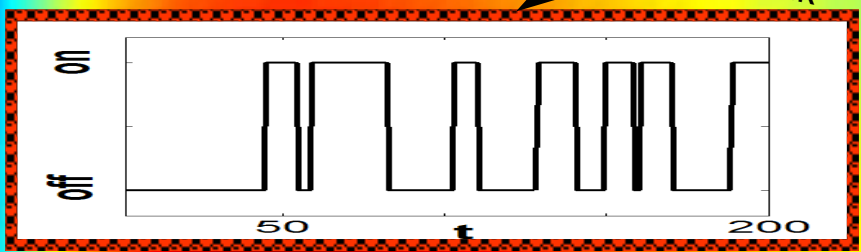
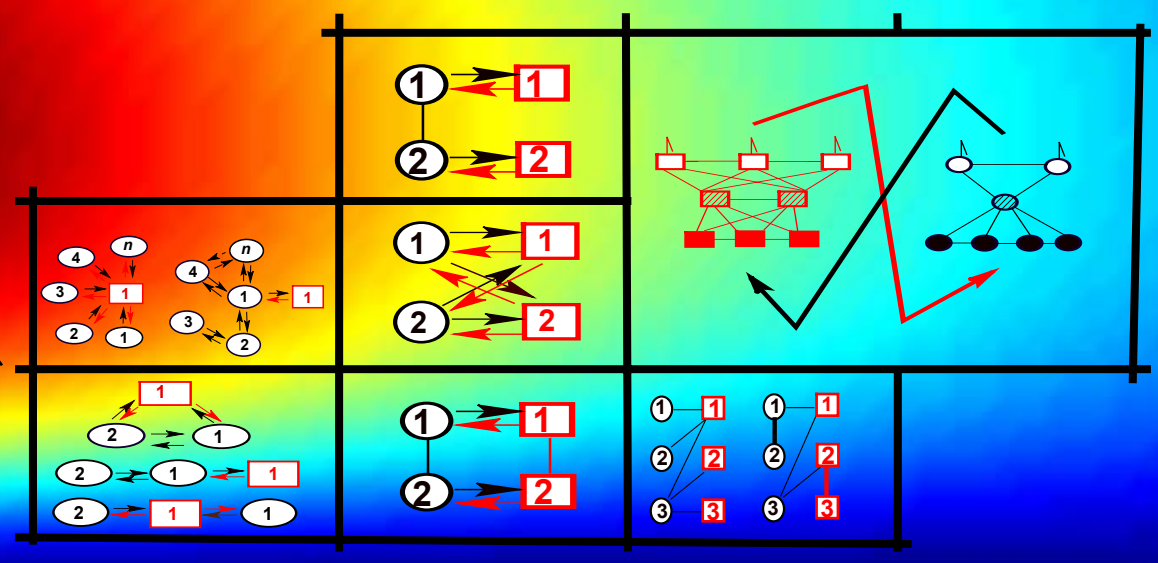